\documentclass[twocolumn,english,aps,prb,twocolum,superscriptaddress,natbib,bibnotes,amsmath,amssymb,floatfix,groupedaddress,footinbib]{revtex4-2}

\usepackage[colorlinks=true,citecolor=blue,linkcolor=magenta]{hyperref}

\usepackage[markup=nocolor, authormarkupposition=left]{changes} 
\usepackage{soul}
\usepackage[utf8]{inputenc}
\usepackage[english]{babel}
\usepackage{amsmath,amsfonts,amssymb}
\usepackage[T1]{fontenc}
\usepackage{url}

\usepackage{amsmath}
\usepackage{siunitx}
\usepackage[version=4]{mhchem}
\usepackage{amsfonts}
\usepackage{amssymb}

\usepackage{epstopdf}
\usepackage{graphicx}
\graphicspath{{./Figures/}}

\usepackage{changes}

\begin{document}

\title{Large-scale photonic computing with nonlinear disordered media}

\author{Hao Wang$^{1,3,\ast}$, 
        Jianqi Hu$^{1,\ast,\dagger}$, 
        Andrea Morandi$^2$, 
        Alfonso Nardi$^2$, 
        Fei Xia$^{1}$,
        Xuanchen Li$^{2}$, 
        Romolo Savo$^{2,4}$, 
        Qiang Liu$^{3}$,  
        Rachel Grange$^{2}$ and Sylvain Gigan$^{1,\ddag}$}
\affiliation{
$^1$Laboratoire Kastler Brossel\char`,{} École Normale Supérieure - Paris Sciences et Lettres (PSL) Research University\char`,{} Sorbonne Université\char`,{} Centre National de la Recherche Scientifique (CNRS)\char`,{} UMR 8552\char`,{} Collège de France\char`,{} 24 rue Lhomond\char`,{} 75005 Paris\char`,{} France.\\
$^2$ETH Zurich\char`,{} Institute for Quantum
Electronics\char`,{} Department of Physics\char`,{} Optical Nanomaterial
Group\char`,{} 8093 Zurich\char`,{} Switzerland.\\
$^3$State Key Laboratory of Precision Space-time Information Sensing Technology\char`,{} Department of Precision Instrument\char`,{} Tsinghua University\char`,{} Beijing 100084\char`,{} China.\\
$^4$Centro Ricerche Enrico Fermi (CREF)\char`,{} Via Panisperna 89a\char`,{} 00184 Rome\char`,{} Italy.\\
}

\maketitle

\noindent\textbf{\noindent
Neural networks find widespread use in scientific and technological applications, yet their implementations in conventional computers have encountered bottlenecks due to ever-expanding computational needs. 
Photonic neuromorphic hardware, which manipulates information and represents data continuously in the optical domain, is one of the promising platforms with potential advantages of massive parallelism, ultralow latency, and reduced energy consumption. While linear photonic neural networks are within reach, photonic computing with large-scale optical nonlinear nodes remains largely unexplored. 
Here, we demonstrate a large-scale, high-performance nonlinear photonic neural system based on a disordered polycrystalline slab composed of lithium niobate nanocrystals. 
Mediated by random quasi-phase-matching and multiple scattering, linear and nonlinear optical speckle features are generated as the interplay between the simultaneous linear random scattering and the second-harmonic generation, defining a complex neural network in which the second-order nonlinearity acts as internal nonlinear activation functions. 
Benchmarked against linear random projection, such nonlinear mapping embedded with rich physical computational operations shows improved performance across a large collection of machine learning tasks in image classification, regression, and graph classification with varying complexity. Demonstrating up to 27,648 input and 3,500 nonlinear output nodes, the combination of optical nonlinearity and random scattering serves as a scalable computing engine for diverse applications. 
}

\section*{Introduction} 

\noindent{From} combinatorial optimization algorithms that exhaustively search solutions of real-world problems to artificial neural networks that support substantial technology revolutions, modern digital computers have been excellent hardware and major workhorse for decades. 
However, they recently appear struggling to keep pace with the booming trend of operation-dense applications by consuming far too much energy, memory, and time for data management, storage, training and inference \cite{wu2022sustainable}. 
A successful large language model takes 1 million GPU hours for training and consumes approximately 0.003 kilowatt-hour electricity per customer query \cite{luccioni2022estimating}. 
The main bottleneck of contemporary digital computers can be attributed to the architecture they are built upon - the von Neumann architecture - where the memory and the central processing unit are physically separated and information is processed sequentially.
A fundamentally different path in computation is to bring the memory or signal to the processing units as close as possible, thereby achieving speed and efficiency advantages.
There has been a continuous drive to develop new neural network accelerators that employ various technologies, from electronics \cite{wang2019benchmarking}, mechanical \cite{wright2022deep} and photonic neuromorphic hardware \cite{wetzstein2020inference,shastri2021photonics} to  memristor \cite{yao2020fully,li2019long} and DNA computing devices \cite{okumura2022nonlinear}. All of these endeavors are in a shared wish to extend brain-inspired information processing to physical systems, for more scalable, environment-friendly, and even capacity-enhanced machine learning (ML) \cite{mehonic2022brain}.

Photonic neural systems are one of the promising platforms for analog computing that show prospects for massive parallelism, ultralow computing latency, and reduced power consumption \cite{wetzstein2020inference,shastri2021photonics}. 
Photonic computing has a rich history of exploration, dating back to the Hopfield network \cite{farhat1985optical} and optical correlators \cite{weaver1966technique} in the past century using either incoherent light intensity or coherent fields. 
Recently, the field of optical computing has been revived following the rapid advancements of deep neural networks, as optics is well suited for the inference tasks of computing owing to the abovementioned advantages. Various distinct matrix-vector multipliers have been implemented in photonic chips for optical inference \cite{shen2017deep,tait2017neuromorphic,feldmann2021parallel,xu202111}. Free-space light propagation through linear diffractive \cite{lin2018all,zhou2021large} or convolutional layers \cite{chang2018hybrid,miscuglio2020massively} has also been leveraged to execute multiplication at a large scale. 
Apart from photonic systems that are isomorphism of digital neural networks, computation is also naturally performed when light scatters through disordered optical media \cite{gigan2022imaging}. Feeding the system with e.g. spatially modulated light, multiple scattering computes random projection - a ubiquitous operation that is broadly used in mathematics and statistics \cite{johnson1984extensions,rahimi2007random}. The optical random projection can be of an extreme large scale, and its random matrix is fixed and physically stored in the medium itself instead of auxiliary memory. Kernel ridge regression can leverage such large-scale optical random projection for ML applications \cite{saade2016random,ohana2020kernel}. They are easy to train with only a readout layer at the output, while leaving the optical part, where most of the computations are performed, untrained. Additionally, optical random projection has shown remarkable capabilities for optical echo state networks \cite{rafayelyan2020large}, Ising machines \cite{pierangeli2021scalable,leonetti2021optical}, randomized linear algebra \cite{hesslow2021photonic}, and training \cite{launay2020hardware}. However, multiple light scattering is mostly linear in nature even under high-power or pulsed illumination. Therefore, independent from the medium's thickness, the process remains equivalent to a single layer of an artificial neural network. Similar to digital neural networks, nonlinear activation functions are indispensable to obtain a deep photonic neural system with appreciable expressive model capacity. 
Several methods to realize nonlinear functions have been developed, from square-law detector nonlinearity \cite{lin2018all}, laser-cooled atoms induced nonlinearity \cite{zuo2019all}, phase-change nonlinearity \cite{feldmann2019all}, to electro-optic, opto-electronic, and opto-electro-optic nonlinearities \cite{tait2019silicon,ashtiani2022chip,wang2023image}. Yet, the vast majority of these nonlinearities are either incoherent or convert the optical waves to other formats, thereby posing challenges for subsequent photonic neural processing. Noticeably, the utilization of coherent optical nonlinearity has recently been proposed \cite{marcucci2020theory,nakajima2021neural,zhou2022nonlinear}, and experimentally demonstrated in nonlinear spatiotemporal evolution of optical pulses in multimode fibers \cite{teugin2021scalable}. Leveraging the third-order ($\chi^{(3)}$) nonlinearity of the silica fiber,  the nonlinear photonic neural system is demonstrated up to a few hundreds of nodes, limited by the number of modes supported by the waveguide. 

Unlike the $\chi^{(3)}$ nonlinearity that is ubiquitous across all material platforms, the second-order ($\chi^{(2)}$) nonlinearity is only present in noncentrosymmetric media. As a lower-order nonlinearity, $\chi^{(2)}$ nonlinear effects are more efficient provided with proper phase matching, hence also interesting for nonlinear optical computation. 
In this paper, we demonstrate a large-scale photonic neural network that combines linear scattering and $\chi^{(2)}$ optical nonlinearity for a wide range of ML applications. 
The core processing unit consists in a disordered polycrystalline lithium niobate (LN) slab assembled from nanocrystals \cite{morandi2022multiple}, which not only is multiply scattering but also generates second-harmonic (SH) light assisted by random quasi-phase-matching \cite{baudrier2004random,savo2020broadband}. The linear scattering of both fundamental-harmonic (FH) and SH waves contributes to fixed random weights, while the optical nonlinearity contributes to activation functions, defining a complex photonic neural network described by a massive third-order scattering tensor \cite{moon2023measuring}. 
We encode the input information in the optical wavefront of the FH and record the speckle intensity of the SH at the output of the LN slab. As such, nonlinear features are extracted from the SH speckle, and are then fed into a simple linear regression layer to solve ML tasks. Meanwhile, the optical processing unit also naturally computes the linear speckle at the FH as the experimental baseline, akin to the optical random projection obtained in linear scattering media \cite{saade2016random}. Compared to linear speckle features together with the simulated random projection, we experimentally verify that the nonlinear features exhibit improved representational capabilities, though showing different gains, across 14 diverse datasets of varying complexity in image classification, univariate and multivariate regression, and graph classification. 
These results unequivocally identify the role of optical nonlinearity in large-scale complex photonic neural networks, with the number of input and nonlinear output nodes up to 27,648 and 3,500, respectively, surpassing the state-of-the-art optical computing hardware. The large dimensionality, wide applicability, and potential high efficiency of our nonlinear optical processing unit extend the current photonic neural systems towards scalable architectures and broader scenarios. 

%%%%%%%%%%%%%%%%%%%%%%%%%%%%%%%%%%%%%%%%%%%%
% FIGURE 1 - principle graph
\begin{figure*}[!htp]
  \centering{
  \includegraphics[width = 0.97\linewidth]{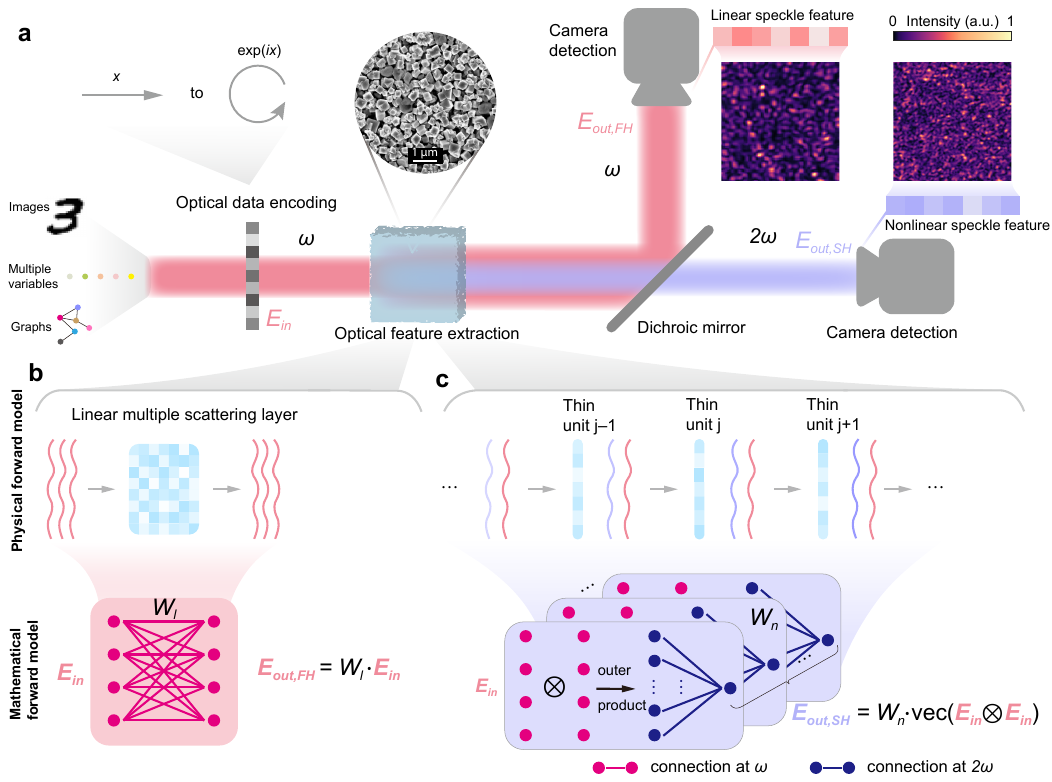}
  } 
    \caption{\noindent\textbf{Optical feature extraction and forward models.} \textbf{a,} The schematic setup for linear and nonlinear feature extraction. 
    A pulsed laser at the FH illuminates a disordered scattering LN slab (a scanning electron microscope image is shown) and generates FH ($\omega$, red) and SH ($2\omega$, violet) speckle features, which are separated by a dichroic mirror and measured by cameras. Input data of various formats are encoded onto the spatial phase of light, and are then transformed to the intensity of the FH and SH speckles (their representative measured speckle patterns are shown, respectively). 
    \textbf{b,} The physical and mathematical forward models of FH speckle features. The linear propagation in a complex medium can be mathematically described as a complex-valued random matrix $W_l \in \mathbb{C}^{N\times M}$, linearly connecting the input field $E_{in} \in \mathbb{C}^{M\times1}$ and the output field $E_{out, F H}\in \mathbb{C}^{N\times1}$. 
    \textbf{c,} The physical and mathematical forward models of SH speckle features. The SH speckle generation process can be dissected by segmenting the LN slab into thin units. In each unit, FH and SH waves experience weak scattering and a portion of the FH is nonlinearly transformed to SH. Overall, the process can be mathematically described by a more complex nonlinear mapping $E_{out, SH} = W_n \cdot {\rm vec}(E_{in}\otimes E_{in})$, where $W_n \in \mathbb{C}^{N\times M^2}$ is a matrix reshaped from a third-order tensor that connects the input field $E_{in} \in \mathbb{C}^{M\times1}$ and nonlinear output field $E_{out, SH}\in \mathbb{C}^{N\times1}$. $\otimes$: outer product; ${\rm vec(~)}$: matrix vectorization. Each output node of $E_{out, SH}$ (violet dot) is nonlinearly connected to input nodes  (one block in $W_n$).
  }
 \label{Figure1}
\end{figure*} 
%%%%%%%%%%%%%%%%%%%%%%%%%%%%%%%%%%%%%%%%%%%%

\section*{Results} 
Figure \ref{Figure1}a shows the schematic setup of our photonic neural system. The entire optoelectronic system comprises data encoding, optical feature extraction, intensity detection and digital readout. First, we encode the input data to the spatial phase profile of the FH pulse at a central wavelength of 800 nm. Noticeably, phase encoding using a liquid-crystal spatial light modulator (SLM) inherently introduces element-wise nonlinearity \cite{antonik2019human}, such that the input data $x$ is transformed to $E_{in} = {\rm exp}(i\pi x)$. Then, the modulated FH wave illuminates a disordered LN slab to generate optical features, thereby projecting input data into a high-dimensional latent space. The feature extraction includes both linear and nonlinear parts: in addition to the linear scattering of the FH wave, the SH at 400 nm is also generated in the LN slab mediated by random quasi-phase-matching and scattered in the meantime \cite{morandi2022multiple}. In the experiment, a dichroic mirror is used to separate them, and additional filtering is required to completely suppress the FH signal in the SH path (see Methods). The intensity of both the linear and nonlinear speckle fields are captured by their respective cameras, forming linear and nonlinear speckle features. For either FH or SH feature extraction, a computationally-cheap digital linear regression layer is trained to perform various ML tasks (see Methods). In this work, we designate the linear and nonlinear features based on their wavelengths, as our primary focus is on the role of optical nonlinearity instead of phase encoding and intensity detection nonlinearity, which are kept identical in both cases for a fair comparison. 

To explain the linear and nonlinear feature computations in detail, we illustrate in Figs. \ref{Figure1}b-c their physical and approximated mathematical forward models. The FH speckle pertains to a classical multiply scattering process in optics, recently exploited as a powerful tool to compute random projection \cite{gigan2022imaging} (see Supplementary Note 1). 
Although the ultrashort FH pulse used in our experiment corresponds to a few spectral correlation bandwidth of the medium \cite{mounaix2016spatiotemporal} (see Supplementary Fig. 3), the linear scattering process can be adequately approximated by a single transmission matrix $W_l \in \mathbb{C}^{N\times M}$ \cite{popoff2010measuring}, connecting the input field $E_{in} \in \mathbb{C}^{M\times 1}$ and output field $E_{out, FH} \in \mathbb{C}^{N\times 1}$ by $E_{out, FH} = W_l\cdot E_{in}$, where $\cdot$ represents the matrix multiplication, and $M$ and $N$ are the dimensions of input and output, respectively. Indeed, the computing results based on our linear speckle features obtained with pulsed excitation are very similar to random projection achieved using continuous-wave (CW) light \cite{saade2016random}.

The formation of the SH speckle involves more intricate processes. To conceptualize such a physical process, we imagine the LN slab is sliced into multiple thin units, with each unit comprising several LN nanocubes (Fig. \ref{Figure1}c). At the input side of a specific unit, we have the FH and generated SH from the previous unit. These incoming FH and SH waves are weakly scattered inside the current unit, described by some fixed and linear connections. Meanwhile, a portion of the FH is nonlinearly transformed into the SH, and add on top on the current SH wave.  As a consequence, the scattered SH fields formed at the camera plane represent the interference of the generated SH from all the units. 
Collectively, the physical interconnection between the input field $E_{in} \in \mathbb{C}^{M\times 1}$ and nonlinear output field $E_{out, FH} \in \mathbb{C}^{N\times 1}$ can be accurately modeled as $E_{out, SH} = W_n \cdot {\rm vec}(E_{in}\otimes E_{in})$, where $\otimes$ and $\rm vec(~)$ denote the outer product operation and the vectorization of a matrix, respectively (see Supplementary Note 1). $W_n \in \mathbb{C}^{N\times M^2}$ is a matrix reshaped from a third-order scattering tensor that fully characterizes the nonlinear scattering process as demonstrated recently \cite{moon2023measuring}, akin to the transmission matrix in the linear scattering case \cite{popoff2010measuring}. Such nonlinear mapping of the input empowers richer physical computations that can be beneficial for a multitude of ML tasks. 
% , encapsulating the complex relation between the input and output optical fields 

%%%%%%%%%%%%%%%%%%%%%%%%%%%%%%%%%%%%%%%%%%%%
 \begin{figure*}[htp]
  \centering{
  \includegraphics[width = 0.9\linewidth]{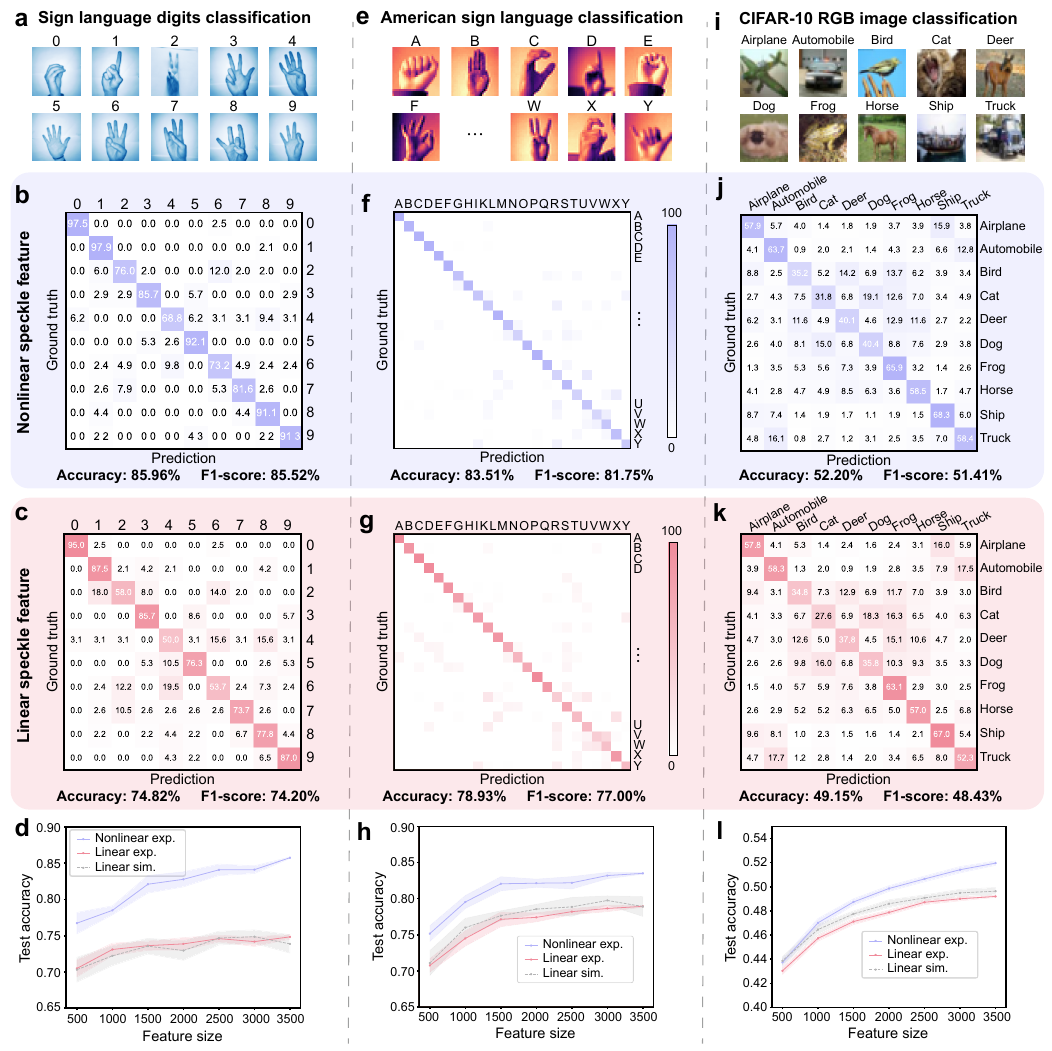}
  } \caption{\noindent\textbf{Speckle features for image classification.} Nonlinear and linear speckle features are used to classify three image datasets: \textbf{a-d} sign language digit (SLD), \textbf{e-h} American sign language (ASL) and \textbf{i-l} CIFAR-10 images. \textbf{a,} 10-class SLD images. \textbf{b,c,} The confusion matrices on the test set obtained by nonlinear (\textbf{b}) and linear (\textbf{c}) speckle features at a feature size of 3,500. \textbf{d,} Comparison of classification accuracies based on the experimental nonlinear feature (violet), experimental linear feature (red), and simulated linear feature (gray) at varying feature sizes. The nonlinear speckle feature outperforms its linear counterpart by a large margin (from 5.4$\%$ to 10.9$\%$). \textbf{e,} 24-category ASL alphabet images. \textbf{f,g,} Same as \textbf{b,c} for ASL alphabet dataset, respectively. \textbf{h,}  Same as \textbf{d}, with test accuracy improvement from 4.0$\%$ to 5.0$\%$ using the nonlinear feature. \textbf{i,} CIFAR-10 RGB images. \textbf{j,k,} Same as \textbf{b,c} for CIFAR-10 dataset, respectively. \textbf{l,} Same as \textbf{d}, with test accuracy improvement from 0.7$\%$ to 2.8$\%$ using the nonlinear feature.
 In \textbf{d,h} and \textbf{l}, the shaded area represents one standard deviation from the mean value calculated from 10 repeated tests.}
  \label{Figure2}
\end{figure*} 
%%%%%%%%%%%%%%%%%%%%%%%%%%%%%%%%%%%%%%%%%%%%

\noindent\textbf{Image classification.} To assess the added value of nonlinearity, we begin by applying our photonic neural system to image classification. 
%Images are Euclidean data and often contain sparse information. 
%With our experimental setup, image features are extracted optically. 
We experimentally extract the optical features and use them to classify the images.
Figure \ref{Figure2} showcases the efficacy of linear and nonlinear optical features for three image classification tasks with varying complexity. 
The first task is to classify 10-class sign language digits (SLD) (Figs. \ref{Figure2}a-d) \cite{mavi2020new}. Sign language is the primary communication language for deaf or hard of hearing people, and recognizing them with machines is helpful for the community. 
Here, we encode the sign language digit (64$\times$64 pixels) to the SLM, with each pixel from the image corresponding to a macropixel at SLM (see Methods and Supplementary Table 1). Then we record the speckle features with cameras and use them for digit classification. 
Figures \ref{Figure2}b-c show the confusion matrices obtained separately by 3,500 nonlinear and linear optical feature nodes, with test accuracies of $85.96\%$ and $74.82\%$, respectively. 
% A remarkable 10.9\% average accuracy improvement is observed using nonlinear speckles at the same feature size of linear speckles. 
In addition to the {\it{accuracy}} metric, we also employ the {\it{F1-score}} to evaluate classification performance \cite{taha2015metrics} (see Methods). For this task, the {\it{F1-scores}} for nonlinear and linear features are 85.53\% and 74.20\%, respectively. 
To validate whether the nonlinear advantage is concrete and universal,  we search the optimal Tikhonov regularization parameters for both linear and nonlinear speckle features at different feature sizes, and also simulate the linear random projection as the benchmark. 
Figure \ref{Figure2}d presents the experimental results of nonlinear and linear speckle features, as well as the simulated random projection. All of them are marked with one standard deviation from 10 repeated trials for each case. It can be seen that the experimental linear features and simulated random projection show very similar classification performance, suggesting that the linear features serve as a fair experimental baseline for ideal random projection. The experimental nonlinear features consistently outperform its linear counterpart at various feature sizes, achieving average accuracy gains ranging from $5.4\%$ to $10.9\%$. 

 Then we proceed to the American sign language (ASL) alphabet dataset (Figs. \ref{Figure2}e-h) \cite{slmnist}, which consists of 27,455 training and 7,172 test images that represent 24 English letters (except J and Z). In our experiment, nonlinear speckles show as more expressive features with a test accuracy of 83.51\%, compared to 78.53\% obtained with linear speckles. An average accuracy improvement of 4.6\% is obtained for feature sizes ranging from 500 to 3,500. Notably, such an accuracy gap may not be easily filled by increasing the dimension of linear features, as it will eventually approach certain kernel limit \cite{saade2016random}. Additionally, we test our photonic neural system for commonly used Modified National Institute of Standards and Technology (MNIST) \cite{MNIST} and FashionMNIST \cite{xiao2017fashion} datasets (see Supplementary Fig. 1). For these tasks, linear optical random projection is already a quite effective approach to extract features, as evident by the classification accuracy compared to relevant works \cite{oguz2022programming,momeni2022electromagnetic,wright2022deep}, but we still observe a performance improvement using nonlinear speckle features.
 
Moreover, we challenge our photonic neural system with more demanding image classification tasks. Figures \ref{Figure2}i-l demonstrate the experimental classification results for the CIFAR-10 (Canadian Institute For Advanced Research) database \cite{krizhevsky2009learning}. Here, the three RGB channels and its generated grayscale channel of each CIFAR-10 image are simply encoded within four neighboring patches in the SLM (see Methods). 
While there may be better encoding strategies available, we manage to attain an accuracy improvement of approximately 3.1\% (52.20\% versus 49.15\%) on 10,000 test images at a feature size of 3,500 (Figs. \ref{Figure2}j-k). 
In addition, we also tackle another challenging image recognition task, the STL-10 database \cite{coates2011analysis} that features a large input dimension of 27,648. The experimental results for classifying STL-10 (see Supplementary Fig. 1) and CIFAR-10 (Fig. \ref{Figure2}l) at various feature sizes show similar behaviors and indicate the significance of incorporating nonlinearity for improved performance.

%%%%%%%%%%%%%%%%%%%%%%%%%%%%%%%%%%%%%%%%%%%%
\begin{figure*}[!htp]
  \centering{
  \includegraphics[width = 0.92\linewidth]{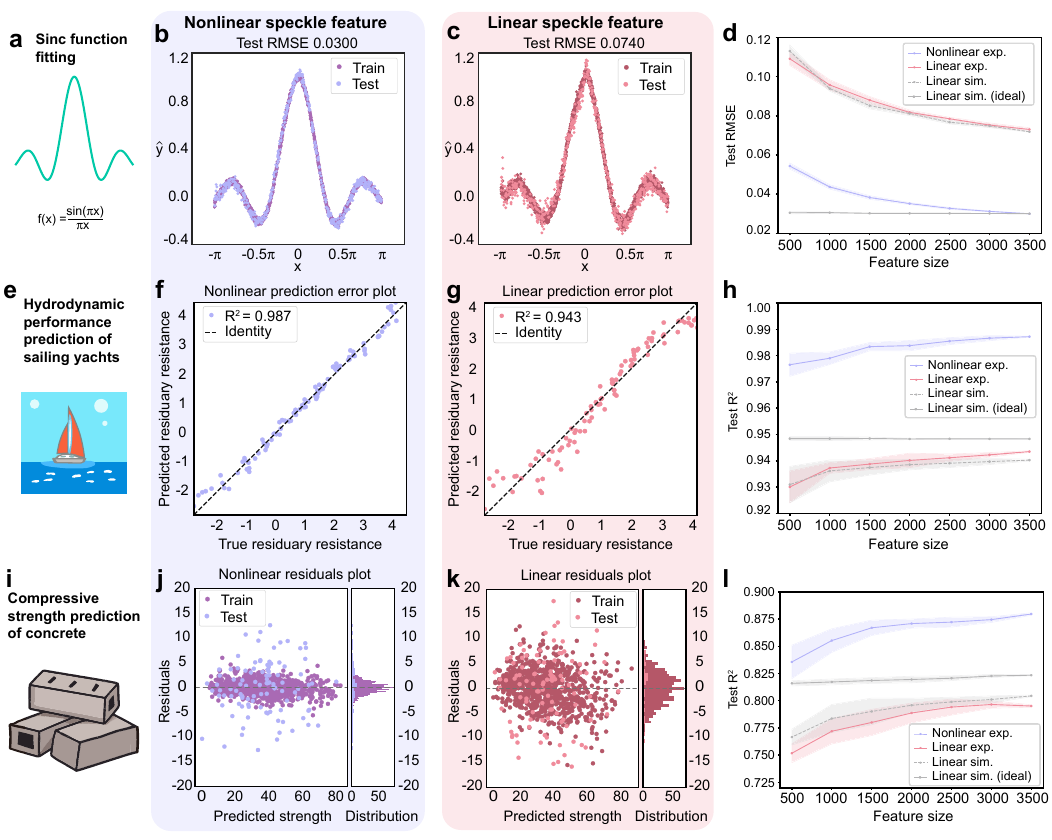}
  } 
    \caption{\noindent\textbf{Speckle features for regression.} Nonlinear and linear speckle features are used for three regression tasks: \textbf{a-d} sinc interpolation, \textbf{e-h} hydrodynamic performance prediction of yachts, and \textbf{i-l} compressive strength prediction of concrete. \textbf{a,} Interpolation of a sinc function $f(x) = \sin(\pi x)/(\pi x)$ for $x \in [-\pi,\pi]$. \textbf{b,c,} The train and test sinc interpolation obtained by nonlinear (\textbf{b}, test RMSE of $0.0300$) and linear (\textbf{c}, test RMSE of $0.0740$) speckle features at a feature size of 3,500. 
    \textbf{d,} Comparison of the test set regression errors based on the experimental nonlinear feature (violet), experimental linear feature (red), and simulated linear features with and without noise (dotted gray and solid gray, respectively) at varying feature sizes. 
    \textbf{e,} Prediction of the hydrodynamic performance of sailing yachts based on their dimensions and velocity (6 input parameters). 
    \textbf{f,g,} Plots of prediction versus ground truth obtained by nonlinear (\textbf{f}, test $R^2$ of $0.987$) and linear (\textbf{g}, test $R^2$ of $0.943$) speckle features at a feature size of 3,500. 
    \textbf{h,} Same as \textbf{d}, but the metric used is the $R^2$. \textbf{i,} Prediction of the concrete compressive strength based on its composition and age (8 input parameters). \textbf{j,k,} The train and test residual error distributions obtained by nonlinear (\textbf{j}) and linear (\textbf{k}) speckle features at a feature size of 3,500.  \textbf{l,} Same as \textbf{h}. In \textbf{d,h} and \textbf{l}, the experimental nonlinear feature (violet) demonstrates improved performance compared to its linear counterpart (red) across all three regression tasks, surpassing even the performance achieved with ideal linear feature (solid gray) in multivariate regressions. 
  }
 \label{Figure3}
\end{figure*} 
%%%%%%%%%%%%%%%%%%%%%%%%%%%%%%%%%%%%%%%%%%%%

\noindent \textbf{Univariate and multivariate regression.}
Our photonic neural system can also be adapted for regression tasks as shown in Fig. \ref{Figure3}. 
Different from the preceding image classifications where the system makes a decision based on the maximum likelihood, regression necessitates directly aligning with precise values. As a result, regression tasks are more vulnerable to the influence of quantization and noise. In our system, both the SLM encoding and camera readout have 8-bit quantization, and the system's signal to noise ratio (SNR) is measured to be around 14 dB, with the noise dominated by the flickering of the SLM, laser fluctuation, and mechanical instability of
the nonlinear scattering sample and optical elements \cite{pierangeli2021scalable}. A canonical regression example is to fit a nonlinear sinc function $y = \sin(\pi x)/(\pi x)$, where $x \in [-\pi,\pi]$ (Figs. \ref{Figure3}a-d). We encode each input variable $x$ to the SLM (see Methods) and record optically generated speckle features. 
In order to fit the corresponding $y$ value, these features must exhibit high nonlinearity regarding $x$ in the given interval. For such univariate regression, we note that the phase encoding nonlinearity plays an important role, as a larger phase encoding range provides higher-order nonlinear terms (see Supplementary Note 2). An encoding range of $[0,2\pi]$ is used in Figs. \ref{Figure3}a-d. Figures \ref{Figure3}b-c show the experimental results of sinc interpolation based on 3,500  nonlinear and linear speckle features, respectively. 
Figure \ref{Figure3}d illustrates their test root mean square errors (RMSEs) at different feature sizes, together with the linear speckle simulation with and without considering the experimental SNR. 
It is evident that using nonlinear speckle features achieves lower RMSEs, owing to the higher-order nonlinear components embedded within them. We also repeat the  experiment above with different phase encoding ranges, and found out that the nonlinear speckle gives lowest RMSE of 0.0069 under an encoding range of $[0, 4\pi]$ (see Supplementary Note 2).

In addition to the univariate regression problem, we also explore two real-world multivariate applications. In Figs. \ref{Figure3}e-h, we predict the hydrodynamic performance of sailing yachts based on their basic hull dimensions and the boat velocity \cite{misc_yacht_hydrodynamics_243}. Figures \ref{Figure3}f-g show the regression results based on nonlinear and linear speckles at a fixed feature size of 3,500. We employ the R-squared ($R^2$) metric here to quantify the regression performance (see Methods), and observe an improved $R^2$ using nonlinear speckle features (0.987 versus 0.943). 
Similarly, in another regression task that predicts the compressive strength of concrete from its composition and age \cite{misc_concrete_compressive_strength_165} (Figs. \ref{Figure3}i-l), the nonlinear speckle gives a residual distribution that is closer to zero, again indicating better performance. Furthermore, for these multivariate regression tasks, the experimental nonlinear features even surpass the ideal linear random features simulated without noise (Figs. \ref{Figure3}h and \ref{Figure3}l). 

%%%%%%%%%%%%%%%%%%%%%%%%%%%%%%%%%%%%%%%%%%%%%
%FIGURE 4 - Sweep for comb and TPM images
\begin{figure*}[!htp]
  \centering{
  \includegraphics[width = 0.97\linewidth]{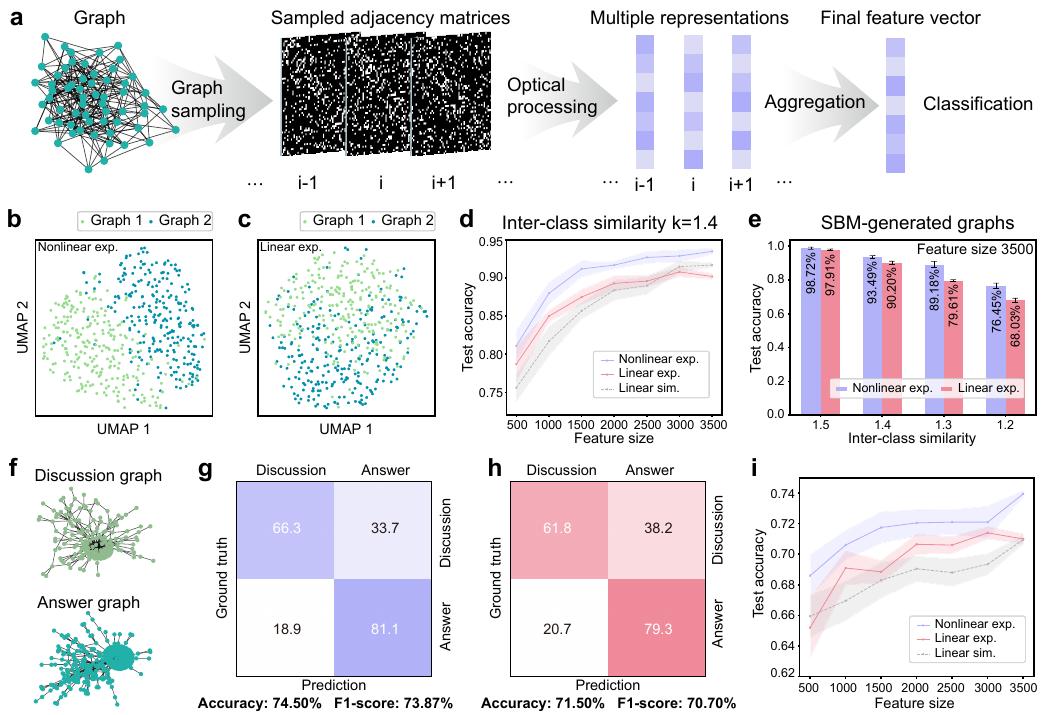}
  } 
    \caption{\noindent\textbf{Speckle features for graph classification.} Nonlinear and linear speckle features are used to classify two graphs datasets: \textbf{b-e} synthetic graphs generated by the stochastic block model (SBM) and \textbf{f-i} Reddit-binary (RB) graphs.
    \textbf{a}, The general graph processing workflow. Multiple adjacency matrices are sampled from a graph, and their corresponding speckle features are optically extracted in our photonic neural system. The average of these  speckle representations is taken as the final feature vector for classification. 
    \textbf{b,c,} Visualizations of SBM-generated graph data (inter-class similarity $k=1.4$) with UMAP obtained by nonlinear (\textbf{b}) and linear (\textbf{c}) speckle features at a feature size of 3,500. 
    \textbf{d,} 
    Comparison of classification accuracies based on the experimental nonlinear feature (violet), experimental linear feature (red), and simulated linear feature (gray) at varying feature sizes. The accuracy gain varies from 2.1$\%$ to 3.7$\%$ for $k=1.4$. \textbf{e,} The bar chart of test accuracies using nonlinear and linear features at varying inter-class similarities.
    \textbf{f,} Illustration of discussion-based and question/answer-based graph samples from the RB dataset. 
    \textbf{g,h,} The confusion matrices on the test set obtained by nonlinear (\textbf{g}) and linear (\textbf{h}) features at a feature size of 3,500. \textbf{i,} Same as \textbf{d}, with test accuracy improvement from 0.7$\%$ to 3.4$\%$ using the nonlinear feature.
  }
 \label{Figure4}
\end{figure*} 
%%%%%%%%%%%%%%%%%%%%%%%%%%%%%%%%%%%%%%%%%%%%%

\vspace{0.1cm}

\noindent\textbf{Graph classification.} Real-world objects can sometimes be more complicated than Euclidean data like images discussed above. Often, they are described by their connections to other things. Such a collection of objects, along with their interrelations, can naturally be represented as a graph with nodes and edges, such as social networks and chemical molecules. Numerous neural network frameworks have been developed to learn representations from non-Euclidean graph data.
From neuromorphic hardware point of view, graph processing has recently been realized in optical linear diffractive units \cite{yan2022all}, linear light scattering \cite{ghanem2021fast}, and electrical analog resistive memory arrays \cite{wang2023echo}. 
In our system, for moderate-sized graphs with similar number of nodes across graphs, we directly encode the adjacency matrix of each graph to the SLM and optically extract representations. 
As shown in Figs. \ref{Figure4}b-e, we test our method on a synthetic dataset composed of two-category graphs generated by the stochastic block model (SBM) \cite{lee2019review}. 
For this dataset, we fix the node number of all graphs constant (60 in this study), and control the difficulty of graph classification via the inter-class similarity $k$ (closer to 1 indicates a higher similarity between the two classes). 
To better understand and visualize the superiority of nonlinear speckle features, we exploit the uniform manifold approximation and projection (UMAP) \cite{mcinnes2018umap} technique to reduce both nonlinear and linear feature dimensions, from 3,500 to 2 (Figs. \ref{Figure4}b-c). Indeed, at a setting of $k = 1.4$, the nonlinear embeddings render better clustering of the two classes, leading to a higher test accuracy (Fig. \ref{Figure4}d). 
Similar accuracy trends are observed with varying inter-class similarities as summarized in Fig. \ref{Figure4}e.

Moreover, we also address a real-world Reddit-Binary (RB) graph database \cite{KKMMN2016} (Figs. \ref{Figure4}f-i). Theses graphs vary significantly in size, from as small as 6 nodes and 4 edges to as large as 3782 nodes and 4071 edges, thus posing challenges for graph data encoding. 
To this end, we develop a generalized graph processing workflow as shown in Fig. \ref{Figure4}a. 
Given a large graph, we first sample multiple subgraphs with a uniform number of nodes and extract the adjacency matrix of each graph. These adjacency matrices are then divided into groups, and each group is subsequently encoded to the SLM altogether. Consequently, multiple subgraph representations are obtained. They are aggregated as final features via an average operation, before being fed into a digital classifier. The detailed parameters used for processing the RB graph dataset are described in the Methods section. Our goal here is to discern between 'discussion-based' and 'question/answer-based' graphs (Fig. \ref{Figure4}f). Again, the nonlinear representations show better classification results and reaffirm the added value of nonlinearity as presented in Figs. \ref{Figure4}g-i.

\vspace{0.1cm}

\section*{Discussion} 
We demonstrate that the synergy of multiple light scattering and SH nonlinearity in our disordered LN slab can extract superior features for a wide variety of ML applications, when compared to its linear FH counterpart (see Supplementary Table 1). 
To unveil the origin of such superiority, we first analyze the nonlinear optical projection based on the mathematical forward model presented in Fig. \ref{Figure1}c. The SLD dataset with the image size of $64^2$ is considered as an example. Different from linear speckle simulation (see Methods), simulating the nonlinear optical mapping would require a massive matrix $W_n$ with $3,500 \times (64^2)^2 = 58,720,256,000$ entries, equivalent to 437.5 GB of memory usage in a digital hardware for an output feature size of 3,500 (see Supplementary Note 3). Matrix-vector multiplication at such an extreme scale becomes computationally-intense and memory-hungry. In our analog system, the computations are optically performed without measuring and storing the matrix, estimated as 18.8 tera floating point operations per second (TFLOPS) albeit the matrix is fixed (see Supplementary Note 3). To examine the role of nonlinearity in a simulation-friendly manner, we consider a much simplified nonlinear optical model if we assume only scalar fields instead of vector fields (see Supplementary Note 1). 
Under such a simplification, the forward model reduces to a two-layer neural network, i.e., $E_{out, SH} = W_{l2}\cdot(W_{l1} \cdot E_{in})^2$, where $(~)^2$ denotes element-wise square function. $W_{l1}$ and $W_{l2}$ are two complex random matrices denoting the linear relations between the input and generation sites at the FH and between the generation sites and output at the SH, respectively. The square nonlinear activation in the hidden layer is a consequence of SH generation, as the generated SH field is proportional to the square of the incoming scalar FH field, i.e., $E(2\omega) \propto E(\omega)^2$. In the simulation, we tune the hidden layer size, which corresponds to the number of SH generation sites in the experiment. 
As we vary the number of hidden neurons from 1,000 to 10,000, the classification accuracies achieved with 500 to 3,500 features range from 74.09\% to 83.29\%, surpassing 70.31\% $\sim$ 74.86\% in the linear simulation case (see Supplementary Note 1 and Supplementary Fig. 6). The simplified nonlinear simulation results also outperform the experimental linear features, yet they fall behind the experimental nonlinear results in Fig. \ref{Figure1}d (76.93\% $\sim$ 85.76\%). This indicates richer computation beyond a two-layer neural network is achieved in the experimental settings. 
Indeed, the simplified simulation does not account for the full second-order nonlinear tensor and randomized orientations and sizes of LN nanocrystals \cite{muller2021modeling} (see Supplementary Note 1). 

The exceptional performance of our photonic neural system stems from its large dimensionality and internal nonlinear activations. The possible scaling of our system is discussed in Supplementary Note 3. Nevertheless, using a table-top femtosecond laser for optical computing in the current study comes with a compromise in energy efficiency. 
Noticeably, sum-frequency speckles have been generated in scattering assemblies of LN nanocrystals with CW illumination \cite{ni2023nonlinear}. This three-wave mixing relies on the same $\chi^{(2)}$ nonlinearity of SH generation. Obtaining CW-pumped SH speckles from our disordered LN platforms is an exciting perspective, as it would dramatically reduce the computation power needed in the system, and the total power consumption in optics can be within tens of watt considering the laser, SLM and camera used for experiments. 
Such an efficient physical transformation performed by the LN slab can also be employed as a building block within a deep neural network \cite{wright2022deep}. 
By incorporating trainable parameters, either from the analog platform or an auxiliary digital hardware, more sophisticated ML tasks can be performed using backpropagation-free training techniques like forward-forward algorithm \cite{hinton2022forward}. While this work randomly draws speckle features from measured camera images, searching 'superior' features via feature selection methods \cite{cai2018feature} could reduce digital computational overhead. 
Besides, exploiting 'hybrid' features by mixing the linear and nonlinear features may provide better performance. 
All-optical and hybrid optoelectronic neural networks have demonstrated potential in various fields such as fiber telecommunication \cite{huang2021silicon}, computational imaging \cite{luo2022computational}, and structured light measurement \cite{wang2023intelligent}. It remains possible to apply our system to address physical problems broadly beyond ML tasks.
Moreover, the use of $\chi^{(2)}$ nonlinear media for computation also opens up new possibilities for quantum or quantum-enhanced computing with entangled photons \cite{leedumrongwatthanakun2020programmable,krastanov2021room}.

In conclusion, we experimentally demonstrate a large-scale, high-performance photonic computing platform that seamlessly integrates light scattering and optical nonlinearity. This innovative nonlinear feature extraction exhibits universal performance enhancements surpassing linear random projection, across 14 diverse machine learning tasks, spanning image classification, regression, and graph classification. 
Given its extensive dimensionality and versatility, we foresee our system as a powerful analog computing platform, potentially serving as a building block of neural networks for a wide range of applications. 

\vspace{0.5cm}

\noindent\textbf{Methods}
\medskip
\begin{footnotesize}
%\begin{methods}

\noindent\textbf{Sample information.} We assemble the slabs starting from crystalline LN nanoparticles that are produced via solvothermal synthesis. The precursors oxides Nb$_2$O$_5$ (H.C. Starck, 99.92\%) and LiOH (Aldrich, 98\%) are dispersed in a mixture of Ethylene Glycol and distilled water. After ultrasonication, the suspension is poured into a PTFE coated stainless steel acid digestion bomb (model PA4748, volume 120 mL, Parr Instrument Company) and hydrothermally treated at 250° C for 70 h. Afterwards, the reaction product is washed with water by centrifugation. This chemical synthesis allows to precisely control the size distribution of the produced nanoparticles. The nanoparticles have sizes from 100 nm to 400 nm, a linear refractive index of n $\approx$ 2.3, and negligible absorption at visible wavelengths \cite{Weis1985,Volk2009}. They have a noncentrosymmetric hexagonal R3c crystal structure that enables SH generation. We assemble the nanoparticles into slabs by drop deposition and solvent evaporation. An aqueous suspension (1 wt \%) of the LN nanoparticles is mixed with a ratio of 60:1 polyvinyl alcohol and deposited over a glass substrate framed with hydrophilic tape. The sample is placed onto a horizontal substrate holder and kept at 0 degrees Celsius for 24 hours. At the end of the process, the nanoparticles are deposited and the water evaporated. The slabs have a subwavelength transport mean free path which results in multiple light scattering within the thickness of 5 $\mu$m (measured by profilometry). A thorough explanation of the fabrication process and further characterization of the linear and nonlinear optical properties of the slabs is reported in Ref. \cite{morandi2022multiple}.

\vspace{0.1cm}

\noindent\textbf{Experimental setup.} The detailed experimental setup is shown in Supplementary Fig. 2. A mode-locked Ti:sapphire laser (MaiTai HP, Spectra-Physics) is used as the pump, producing 80 MHz repetition-rate, 100 femtosecond laser pulses at a central wavelength of 800 nm. The laser beam goes through a half-wave plate (HWP) and a polarizing beam splitter (PBS), in order to align the beam polarization to the working-axis of the SLM as well as to adjust the light intensity. A spectrometer (Ocean optics, HR4000) collects a small portion of light to monitor the emission spectrum of the laser. The laser beam is expanded by a $4-f$ system via two lenses (L1, $f=12$ mm; L2, $f = 100$ mm) and then illuminates a reflective phase-only SLM (HSP512L-1064, Meadowlarks). Datasets are encoded to the spatial wavefront of the light via the SLM, which is relayed by another $4-f$ system (L3, $f = 150$ mm; L4, $f = 150$ mm) to the back focal plane of an objective lens (Obj. 1, Plan N $20\times 0.4$, Olympus). As such, the pump is delivered to the scattering LN slab via Obj. 1, and the scattered pump and the generated SH light are collected by an identical objective lens (Obj. 2). 
A dichroic mirror (DMSP650r, Thorlabs) is used to separate the FH and SH components. Both FH and SH speckles are captured by their respective cameras (Manta, G-046, Allied Vision) via tube lenses (L5, $f = 100$ mm and L6, $f=150$ mm). In the FH path, a linear polarizer (LP1, LPVIS100-MP, Thorlabs) and neutral density filters are used to select the beam polarization vertical to the incident light and attenuate the FH intensity, respectively. 
While in the SH path, a linear polarizer (LP2, LPUV100-MP2, Thorlabs) is utilized to select linear polarized SH, and a bandpass filter (FBH400-40, Thorlabs) is also employed to completely suppress the residual pump in the SH path. In the experiment, the FH and SH speckles are measured separately due to the dispersion of the collecting objective lens. 

\vspace{0.1cm}

\noindent\textbf{Data processing in experiments and simulations.}
We employ a phase-only SLM to encode input data into the optical fields, with a phase encoding range of $[0, \pi$] unless otherwise specified. We use the central region of the SLM (around 400$\times$400 pixels) for data encoding, while the periphery pixels are blocked by the aperture of the objective lens. Different macropixel sizes are used depending on different data dimensions (see Supplementary Table 1). In cases the central region is not fully used, the unmodulated pixels act as a static bias. 
For grayscale image datasets like SLD, ASL, MNIST and FashionMNIST, we directly display each input image on the SLM. 
While for color image datasets like CIFAR-10 and STL-10, we concatenate three channels of each image (red, green and blue) together with its grayscale version as four blocks of a single composite image. As for univariate regression, the input variable cannot be directly encoded on the SLM, because a uniform phase change will not alter the generated speckle features. Instead, we employ a fixed, small-scale linear random matrix (30$\times$30), and encode the product of the input variable and the matrix to the SLM, as has been implemented previously \cite{teugin2021scalable}. Different phase encoding ranges are investigated (see Supplementary Note 2), and this is the only case we explore encoding range other than $[0, \pi$]. In contrast to univariate regression, input variables are directly encoded on the SLM in multivariate regression with proper macropixel sizes. 
For SBM generated and Reddit-binary graph datasets, we encode the adjacency matrix of the graph to the SLM. In this study, we focus solely on undirected graphs and binary adjacency matrices with only connection information, though the proposed system is not limited to such cases.
All SBM graphs consist of 60 nodes therefore the dimensions of their corresponding adjacency matrices are $60\times60$. Those adjacency matrices are directly displayed on the SLM. For RB database, we first use the random walk method \cite{leskovec2006sampling} to sample each original graph with a sample size of 6 vertices for 2,000 times. Subsequently, we group the 2,000 sampled adjacency matrices into 20 distinctive phase masks, each comprising 100 matrices. These phase masks are loaded sequentially to the SLM to derive 20 speckle representations, which are then averaged to form as the final graph representation used for following digital readout.

In order to extract optical features, we downsample the measured speckle images with a stride size matching the speckle grain size. The latter is estimated by speckle auto-correlation. Therefore, distinctive and roughly independent optical features are obtained and are further flatten as a feature vector (feature size $\sim$ 3500 for both FH and SH speckles). Feature vectors with smaller feature sizes are randomly drawn from the complete feature vectors.
 
The optically generated feature vectors are then fed into a digital readout layer for ML tasks. It performs a simple linear regression trained by the Tikhonov regularization method. 
To do so, we organize the $N$-dimensional feature vectors of $K$ samples into a matrix $X$. The objective is to find a linear combination of the feature vector elements that can best predict each label (grouped in a vector $y$). In other words, we aim to adjust the weight matrix $W_{out}$ to minimize the $\ell_2$ norm of the residual, i.e., $\Vert X \cdot W_{out} - y\Vert_2^2$. 
To penalize large weight values, we add an additional regularization term $\beta\Vert W_{out}\Vert_2^2$ to the objective function. 
This gives the analytical solution of the readout matrix as $\hat{W}_{out}=(X^T\cdot X+\beta I)^{-1}\cdot X^T \cdot y$.
An optimal regularization parameter $\beta$ can enhance the model's generalization ability or reduce the risk of overfitting, particularly when $K < N$. As such, we search the best $\beta$ in all cases for fair comparisons. 

To simulate the computing performance of speckle features, we initialize random Gaussian independent and identically distributed complex matrices, whose dimensions are determined by the input data size and output feature size. For linear speckle simulation, we employ a single random matrix in the digital model, rather than multiple of them corresponding to a multi-spectral transmission matrix \cite{mounaix2016spatiotemporal}. This is valid as we observe the computing performance in these two cases are very similar, while a single random matrix provides a straightforward baseline same as linear random projection using CW light. Compared to image and graph classification tasks, we notice that the regression performance is more sensitive to quantization and noise. Therefore, as shown in Fig. \ref{Figure3} of the main text, we incorporate the quantization and SNR when simulating the regression tasks. The effect of SNR is emulated by adding white Gaussian noise to the speckle intensity.
Throughout all experimental data analysis and simulations, we use both MATLAB software (MathWorks Inc.) and PyCharm software (JetBrains Inc.) on two computers: one equipped with an Intel(R) Core(TM) i7-6700 CPU and 32 GB RAM and another with an AMD EPYC 7351P CPU, two NVIDIA GeForce RTX 2080 Ti GPUs, and 64GB RAM.

\vspace{0.1cm}

\noindent\textbf{Performance evaluation metrics and visualization techniques.} 
Here we describe metrics and visualization methods used in this work for ML data analysis and evaluation. In classification tasks, \textit{confusion matrix} provides a comprehensive summary of classification model performance, showing the number of \textit{true positive (TP)}, \textit{true negative (TN)}, \textit{false positive (FP)}, and \textit{false negative (FN)} predictions. In this work, each element of the confusion matrix indicates the percentage of predictions relative to the total number of instances in that category.  \textit{Accuracy} quantifies the overall correctness of a classification model, expressing the proportion of correct predictions over the total number of predictions, i.e., \textit{accuracy} $=\frac{TP+TN}{TP+FP+TN+FN}$. The \textit{F1-score} combines \textit{precision} $=\frac{TP}{TP+FP}$ and \textit{recall}  $= \frac{TP}{TP+FN}$ and is defined as the harmonic mean of \textit{precision} and \textit{recall}, i.e., \textit{F1-score} $=\frac{2 \frac{TP}{TP+FP} \frac{TP}{TP+FN}} {\frac{TP}{TP+FP} + \frac{TP}{TP+FN}} = \frac{2TP}{2TP+FP+FN} $. It offers a balanced measure of model performance,  especially when dealing with imbalanced datasets.

For regression tasks, $RMSE=\sqrt{\sum_{i=1}^N \frac{(\hat{y}_i - y_i)^2}{N}}$ is a widely used metric to quantify the average difference between prediction $\boldsymbol{\hat{y}}$ and ground truth $\boldsymbol{y}$. Similarly, $R^2=1-\frac{\sum_{i=1}^N (\hat{y}_i - y_i)^2}{\sum_{i=1}^N (\bar{y} - y_i)^2}$ measures how well the model's prediction fits the actual data, with a higher value indicating a better fit. Error plots visually display the relationship between model predictions and actual targets. By comparing this plot against the 45-degree line ($\hat{y} = y$, where the prediction perfectly matches the ground truth), it facilitates a clear visualization of a model's variance in a given task. Residual plots graphically represent the differences between ground truths and predicted values, giving insights into model accuracy and potential patterns in the errors. The residual plot from a well-fitted regression exhibits a dense cluster of points near the origin, accompanied by a sparse and unbiased distribution away from it.

UMAP is a dimension reduction technique widely employed for visualizing high-dimensional data in a lower-dimensional space, while preserving important relationships and structures of data \cite{mcinnes2018umap}. In brevity, UMAP initially constructs a weighted, high-dimensional graph, wherein each vertex corresponds to a feature vector and each edge weight denotes the likelihood that two vertices are connected. Then UMAP refines a low-dimensional layout to make it as similar to the initial graph as possible.

\vspace{0.1cm}

\noindent \textbf{Acknowledgements}:
This work was supported by Swiss National Science Foundation (SNF) projects LION and APIC (TMCG-2$\_$213713), ERC SMARTIES and Institut Universitaire de France. H.W. acknowledges support from China Scholarship Council. J. H acknowledges SNF fellowship (P2ELP2$\_$199825). R.S. acknowledges support from European Union - NextGenerationEU, project Comp-SECOONDO (MSCA$\_$0000079).

\vspace{0.1cm}

\noindent \textbf{Data Availability Statement}: 
 The code and data used to produce the results within this work are available at \texttt{https://doi.org/10.5281/zenodo.8386366}.\end{footnotesize}

\renewcommand{\bibpreamble}{
$^\ast$These authors contributed equally to this work.\\
$^\dagger${Corresponding author: \textcolor{magenta}{jianqi.hu@lkb.ens.fr}}\\
$^\ddag${Corresponding author: \textcolor{magenta}{sylvain.gigan@lkb.ens.fr}}
}

\bibliographystyle{naturemag}
\bibliography{ref}

\end{document}

\title{Supplementary information for:\\ Large-scale photonic computing with nonlinear disordered media}

\author{Hao Wang$^{1,3,\ast}$, 
        Jianqi Hu$^{1,\ast,\dagger}$, 
        Andrea Morandi$^2$, 
        Alfonso Nardi$^2$, 
        Fei Xia$^{1}$,
        Xuanchen Li$^{2}$, 
        Romolo Savo$^{2,4}$, 
        Qiang Liu$^{3}$,  
        Rachel Grange$^{2}$ and Sylvain Gigan$^{1,\ddag}$
        }
\affiliation{
$^1$Laboratoire Kastler Brossel\char`,{} École Normale Supérieure - Paris Sciences et Lettres (PSL) Research University\char`,{} Sorbonne Université\char`,{} Centre National de la Recherche Scientifique (CNRS)\char`,{} UMR 8552\char`,{} Collège de France\char`,{} 24 rue Lhomond\char`,{} 75005 Paris\char`,{} France.\\
$^2$ETH Zurich\char`,{} Institute for Quantum
Electronics\char`,{} Department of Physics\char`,{} Optical Nanomaterial
Group\char`,{} 8093 Zurich\char`,{} Switzerland.\\
$^3$State Key Laboratory of Precision Space-time Information Sensing Technology\char`,{} Department of Precision Instrument\char`,{} Tsinghua University\char`,{} Beijing 100084\char`,{} China.\\
$^4$Centro Ricerche Enrico Fermi (CREF)\char`,{} Via Panisperna 89a\char`,{} 00184 Rome\char`,{} Italy.\\
}
\maketitle

\noindent{\textbf{\large{Contents}}}

\noindent{\textbf{Supplementary Notes:}}\\
\noindent{1. Linear and nonlinear scattering processes in disordered media}\\
2. Computing performance analysis\\
3. Phase encoding nonlinearity\\
4. System stability

\noindent{\textbf{Supplementary Figures:}}\\
\noindent{Figure S1. Extended image classification results}\\
Figure S2. Experimental setup\\
Figure S3. Linear speckle spectrum correlation\\
Figure S4. Sinc function interpolation with different phase encoding ranges\\
Figure S5. Simulation of SLD image classification with a simplified nonlinear forward model\\
Figure S6. Experimental system stability

\noindent{\textbf{Supplementary Tables:}}\\
\noindent{Table S1. Summary of experimental results}\\

%%%%%%%%%%%%%%%%%%%%%
\section*{\textbf{Supplementary Note 1. Linear and nonlinear scattering processes in disordered media}}

\noindent{In} this section, we provide the relevant background of linear and nonlinear scattering processes in disordered media. 

\subsection{Linear scattering in disordered media.} When a coherent light illuminates a complex and inhomogeneous medium, it undergoes scattering. Such a process involves numerous scattering events that result in a random speckle pattern at the output. 
% This pattern appears random due to the complex interference of all scattering paths and multitude of scattering events. 
% The statistical properties of the speckle fields are well-defined and serve as a unique signature of the specific disordered medium. 
Despite the complexity of the process, light propagation through the scattering media remains a linear process. That said, the output optical field, flattened  as $E_{out, FH} \in \mathbb{C}^{N\times1}$, can be expressed as the product of a transmission matrix (TM) $W_l \in \mathbb{C}^{N\times M}$ and the incident optical field $E_{in} \in \mathbb{C}^{M\times1}$,  i.e., $E_{out,FH} = W_l \cdot E_{in}$. 
Here, the TM describes the optical input-output relation between the SLM and camera. 
When a thick, multiple scattering medium is used, the TM becomes a dense, complex-valued, random matrix with entries following an independent and identically distributed complex Gaussian distribution \cite{gigan2022imaging}. Given that commercial SLMs and cameras typically have millions of pixels, the corresponding TM can be of extreme dimensions ($\sim 10^{6} \times 10^{6}$). For optical computing, we can exploit the high dimensionality of the TM to generate random features of input data, without the need to measure the TM. This is well suited for some machine learning algorithms that do not require explicit form of computations, but instead just information mixing or data domain mapping.
For example, kernel ridge regression based on random projection \cite{saade2016random} (also the linear part of this work) and reservoir computing \cite{rafayelyan2020large} have been demonstrated with a
large and fixed random matrix computed by the optical linear scattering process.

\subsection{Second-harmonic generation in nonlinear disordered media.} SH generation is a nonlinear optical process that doubles the frequency of the input light. In essence, two photons of the same optical frequency (denoted as $E(\omega)$) are up-converted to a photon of twice the frequency (denoted as $E(2\omega)$) in a non-centrosymmetric nonlinear media. Generally, the second-order nonlinear polarizations $\boldsymbol{P}(2\omega)=(P_x(2\omega),P_y(2\omega),P_z(2\omega))$ of a $\chi^{(2)}$ nonlinear crystal in the crystal frame induced by the incoming light $\boldsymbol{E}(\omega)=(E_x(\omega), E_y(\omega), E_z(\omega))$ is described by \cite{boyd2008nonlinear}:
\begin{equation}\label{eq1}
    \begin{pmatrix}
    P_x(2\omega)\\
    P_y(2\omega)\\
    P_z(2\omega)\end{pmatrix}=2\epsilon_0\begin{pmatrix}
        d_{11} & d_{12} & d_{13} & d_{14} & d_{15} & d_{16}\\
        d_{21} & d_{22} & d_{23} & d_{24} & d_{25} & d_{26}\\
        d_{31} & d_{32} & d_{33} & d_{34} & d_{35} & d_{36}
    \end{pmatrix}
    \begin{pmatrix}
        E_{x}(\omega)^2\\
        E_{y}(\omega)^2\\
        E_{z}(\omega)^2\\
        2E_{y}(\omega)E_{z}(\omega)\\
        2E_{x}(\omega)E_{z}(\omega)\\
        2E_{x}(\omega)E_{y}(\omega)\\
    \end{pmatrix}
\end{equation}
where $\epsilon_0$ is the vacuum permittivity and $d_{il} (i=1,2,3, l= 1,2,3,4,5,6)$ form a $3\times6$ matrix contracted from a full nonlinear susceptibility tensor. For the LN material used in this work, the nonzero elements of the matrix are $d_{15} = d_{24} = d_{31} = d_{32}$, $d_{16} = d_{21} = -d_{22}$, and $d_{33}$ \cite{boyd2008nonlinear}. The generated SH field $\boldsymbol{E}(2\omega)$ is proportional to the nonlinear polarization $\boldsymbol{P}(2\omega)$. In our experiment, the LN slab used is not only nonlinear but also multiple scattering for both FH and SH waves. 
After incorporating the linear scattering of FH before the generation and SH after the generation, as well as the local coordinate transformations in the LN crystals \cite{muller2021modeling}, the SH speckle field at the camera plane $E_{out}(2\omega) \in \mathbb{C}^{N\times1}$ is connected to the FH at the input plane $E_{in}(\omega) \in \mathbb{C}^{M\times1}$ by the relation \cite{moon2023measuring}:
\begin{equation}\label{eq2}
    E_{n}(2\omega)=\Sigma_{j,k}^{M} W_{n jk}E_j(\omega) E_k(\omega)
\end{equation}  
where $E_{n}(2\omega)$ represents one output node of the SH field with $n = 1,2,...,N$ depending nonlinearly on the input modes of $j,k = 1, 2, ..., M$. The complex coefficient $W_{njk}$ encapsulates linear light scattering, transformation in local crystal coordinates, and the LN nonlinear susceptibilities as described by Eq. \eqref{eq1}. As such, a scattering tensor with the dimension of $M\times M \times N$ is needed to accurately describe the physical forward process. In the main text, the Eq. \eqref{eq2} above is rewritten as: \begin{equation}\label{eq3}
    E_{out,SH} = W_n \cdot {\rm vec}(E_{in}\otimes E_{in})
\end{equation}
where $\otimes$ is the outer product operation, $\rm vec(~)$ represents the vectorization of a matrix. $W_n \in \mathbb{C}^{N\times M^2}$ is a matrix reshaped from the third-order tensor. Although the entries of this matrix may not be fully independent, for example, $W_{njk} = W_{nkj}$ from the commutative property, it offers rich mathematical operations for photonic computing.

In addition, to understand the added value of incorporating nonlinearity in a random neural network, we simplify our physical model as a ‘two-layer neural network', i.e., $E_{out,SH} = W_{l2} \cdot (W_{l1} \cdot E_{in})^2$, with the simplification of the vector field in Eq. \eqref{eq1} as a scalar. This model also does not consider the pump intensity variation within the sample due to the diffusion of the FH, and random sizes of LN nanocrystals.
The simplified optical process can be decomposed into three sequential steps: 1) linear scattering of FH, 2) SH generation from scalar FH and 3) linear scattering of SH. 
More specifically, we use a transmission matrix of FH ($W_{l1} \in \mathbb{C}^{H\times M}$) to describe the random linear scattering of $E_{in}(\omega) \in \mathbb{C}^{M\times1}$ into the FH fields just before SH generation, denoted by $E_{hidden}(\omega) = W_{l1} \cdot E_{in}(\omega) \in \mathbb{C}^{H\times1}$, where $H$ corresponds to the number of the SH generation sites.  
Subsequently, the SH field is generated with element-wise square nonlinearity, i.e., $E_{hidden}(2\omega)=E_{hidden}(\omega)^ 2 \in \mathbb{C}^{H\times 1}$. 
Then the linear random scattering of SH, described by a second transmission matrix ($W_{l2} \in \mathbb{C}^{N\times H}$), gives the SH field at the camera plane as $E_{out}(2\omega) = W_{l2} \cdot E_{hidden}(2\omega) \in \mathbb{C}^{N\times1}$. Although this toy simulation is a simplified version of the experimental physical process, it helps us visualize the superiority of our nonlinear optical projection (see Supplementary Fig. 5).

\section*{\textbf{Supplementary Note 2. Phase encoding nonlinearity}}
In our experiment, we use a phase-only SLM to encode input data. This naturally introduces a nonlinear operation, as the data is now transformed to the argument of complex numbers in the unit circle. 
This is particularly evident for univariate regression tasks, such as interpolation of a sinc function as demonstrated in Figs. 3a-d in the main text. For the input data $x \in [-\pi, \pi]$, we first linearly translate and scale the data, i.e., $h(x)=\frac{\alpha}{2}(x+\pi)$, such that the transformed data is within the encoding range of $[0,\alpha\pi]$. The actual phase applied to the SLM is $h(x)$ after modulo $2\pi$, which is the same as loading $h(x)$ due to the $2\pi$ periodicity in the phase space. Here, $\alpha$ controls the degree of phase encoding nonlinearity, as the phase encoding can be Taylor-expaned as  $e^{i\frac{\alpha}{2}(x+\pi)} = 1 + i\frac{\alpha}{2}(x+\pi) + (i\frac{\alpha}{2}(x+\pi))^2/2 + \cdots$. 
To investigate such nonlinearity in detail, we vary the encoding range parameter $\alpha \in \{1, 2, 3, 4, 5, 6\}$ in the experiment and record the nonlinear speckle features. The interpolation results are shown in Supplementary Fig. 4. It can be seen that the optimal interpolation (lowest RMSE) is obtained when $\alpha=4$, while the highest RMSE occurs at $\alpha=1$. For $\alpha=1$, the generated SH features are not sufficiently nonlinear to fit the nonlinear sinc function. While towards a very large encoding range ($\alpha=6$), it seem the features are too much nonlinear such that the interpolation performance tends to degrade. In Fig. 3 of the main text, we show the interpolation results for $\alpha=2$, which already exhibits a good fit for this task. 
For all the other datasets in this work, we use $\alpha=1$ which corresponds to a phase encoding range of $[0,\pi]$. This is valid because the highest and lowest values within the original data are mapped to farthest points in the unit circle after phase encoding. 

\section*{\textbf{Supplementary Note 3. Computing performance analysis}} 
The proposed computing hardware harnesses the complexity of a physical system to achieve efficient computations. While it is not straightforward to precisely identifying the actual useful computational operations, we first estimate the number of weights and operations in our photonic hardware based on Eq. \eqref{eq3}. 
As described in the Discussion section of the main text, an extra scale matrix $W_n \in \mathbb{C}^{3,500 \times (4,096)^2}$ is needed to establish the input-ouput relation for the SLD dataset. If we digitalize each entry of the matrix as an 8-byte double-precision floating-point value in a digital hardware, it requires about 437.5 GB of random-access memory to store $W_n$ before actual computation happens. Such a storage demand, however, remains viable exclusively for advanced computer systems. 
Following this description, we are able to further examine the overall number of computational operations in our system. 
In order to benchmark against digital systems, we break down each complex operation into constituent real operations \cite{zhou2021large}. Specifically, a complex multiplication can be decomposed into 4 real multiplications and 2 real summations, while a complex summation entails 2 real summations. 
In the most general sense, there are $NM^2$ complex multiplications and $N(M^2-1)$ complex summations for the multiplication of Eq. \eqref{eq3} while it requires $M(M+1)/2$ complex multiplications to construct the outer product of $E_{in}$ and $E_{in}$. The number of operations associated with phase encoding and intensity detection are omitted. With regards to the digital readout layer, the number of operations can be calculated in the same way. For a final output category/dimension of $K$ of the readout layer, there are $KN$ real multiplications and $K(N-1)$ real summations. 
To sum up, the total number of real operations implemented optically and digitally are $R_o \approx 6M(M+1)/2 + 6NM^2+2N(M^2-1) \approx 8NM^2+3M^2$ and $R_d=2KN-K$, respectively. As such, for the SLD dataset, the number of optical operations ($R_o\approx 4.7\times10^{11}$) is many orders of magnitudes higher than that of electronic operations ($R_d \approx 7.0\times10^4$). 
Since the current system operates at a frame rate of 40 fps, we estimate that the optical system's computation performance is 18.8 tera floating point operations per second (TFLOPS). 
For comparison, advanced commercial GPUs, such as the NVIDIA V100 TENSOR CORE, achieve 7 TFLOPS in terms of double-precision performance. 

While the optically implemented computations with our system are undeniably rich and massive, they remain fixed without programmability in the current work,  thereby not possible to ‘substitute' a commercial general-purpose GPU. On the other hand, we train a convolutional neural network (CNN) in a digital computer that performs classification accuracy on par with our experimental results, allowing us to indirectly assess the computational capabilities of our system \cite{oguz2022programming}. We still use the SLD dataset as an example and we train a classic LeNet-5 \cite{lecun1998gradient} to classify them.  
%When the nonlinear speckle feature size is set to 3,500, the average test accuracy from 10 repeated experiments is 85.76\%. 
To accommodate the image size of the database, we make slight adjustments to the digital network architecture. We use the cross-entropy loss function and the Adam optimizer \cite{kingma2014adam} to train such a network. After 100 epochs of training, the model converges well and we achieve a test accuracy of 86.41\%. These results suggest that the proposed optoelectronic neural network (experimental test accuracy of 85.76\%) is comparable to (slightly lower than) the digital LeNet-5 in this particular task. The total parameters of this CNN is approximately 333,077, estimated by a public neural network analyzer \cite{torchstat}. 
In contrast, the digital readout of our system comprises only 35,000 trainable parameters, representing a mere one-tenth of the digital CNN's parameters. 
Moreover, the trainable parameters in our system are determined by the analytical Tikhonov regularization method, which is much easier compared to error backpropagation in deep learning. In addition, the computation weights of our core processing unit can be accessed instantly as light passes through the slab.  As such, the computing time in our system only scales linearly with the data dimension owing to the electronic overhead \cite{rafayelyan2020large}, in contrast to polynomial-order scaling in von Neumann hardware. 

There are still opportunities to further scale up the current optical neural system. In our experiment, we demonstrate large-scale input nodes and nonlinear output nodes up to 27,648 (STL-10 database, 96$\times$96$\times$3) and 3,500,  respectively.
In principle these numbers can be readily increased to 65,526 input nodes (256$\times$256, considering a macro-pixel of 2$\times$2) and 452,400 (780$\times$580) output nodes using full screens of the encoding SLM and detection camera, thus showcasing the system's potential for handling even larger datasets and higher-dimensional computations \cite{kaplan2020scaling}. 
Our setup utilizes an SLM capable of operating at several hundred frames per second, while the CCD camera's maximum frame rate at full resolution is 67.5 fps, resulting in an overall system frame rate of 40 fps. 
We could upgrade the system by using a camera with a higher frame rate, and replacing the SLM with a digital micromirror device operating at kilohertz speed. 
Moreover, it is possible to implement advanced pulse shaping techniques to further push the system's limits. For example, instead of encoding data frame by frame, multiple inputs may be encoded simultaneously in one frame, along with optical beam steering to sequentially process them. These innovations hold the potential to boost the system's computational speed and performance. 

\section*{\textbf{Supplementary Note 4. System stability}}
System stability is crucial for reliable and consistent computations for any task. During the experiments, we maintain a stable ambient environment with the help of an air conditioner in the lab and two shielding cages to cover the setup and the nonlinear scattering sample, respectively. Supplementary Fig. 6 shows the measured system stability of both nonlinear and linear paths using the femtosecond laser. Specifically, we measure the correlation between the first speckle image and subsequent speckle images while keeping the SLM pattern unchanged. 
It can be seen that the nonlinear path exhibits slightly lower stability compared to the linear path, which is reasonable as the nonlinear process can be more vulnerable to external perturbations \cite{samanta2022speckle}. This level of stability is sufficient to measure an entire ML dataset up to 70,000 samples within half an hour at a frame rate of 40 fps.  

\clearpage
\begin{figure*}[!hp]
  \renewcommand{\figurename}{Supplementary Figure}
  \centering{
  \includegraphics[width = 1.0\linewidth]{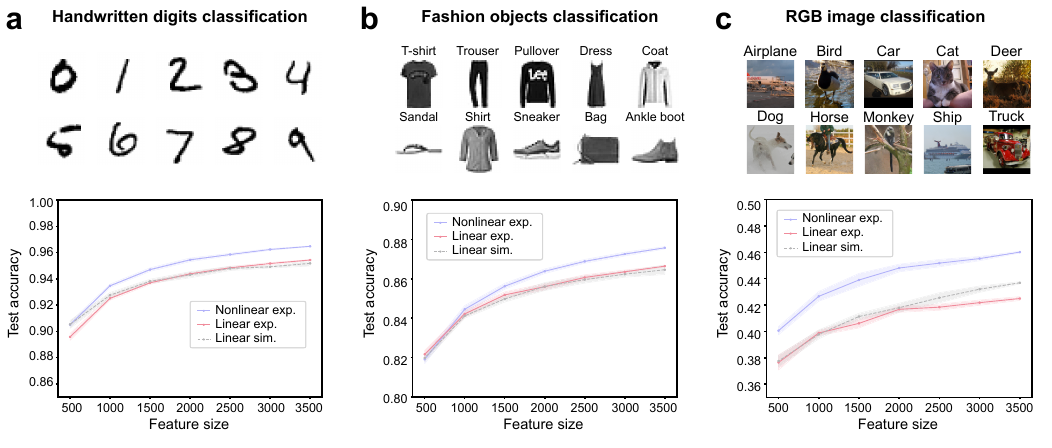}
  } 
  \caption{\noindent\textbf{Extended image classification results.} Top: 10 representative images from \textbf{a} MNIST, \textbf{b} FashionMNIST, and \textbf{c} STL-10 (RGB) image datasets. Bottom: their corresponding test accuracies based on experimental nonlinear features (violet), experimental linear features (red), and simulated linear random projection (grey). The maximum performance gain from nonlinear speckles is observed as 3.5$\%$ in STL-10 case.}
 \label{FigureS1}
\end{figure*}

\begin{figure*}[!hbp]
  \renewcommand{\figurename}{Supplementary Figure}
  \centering{
  \includegraphics[width = 0.8\linewidth]{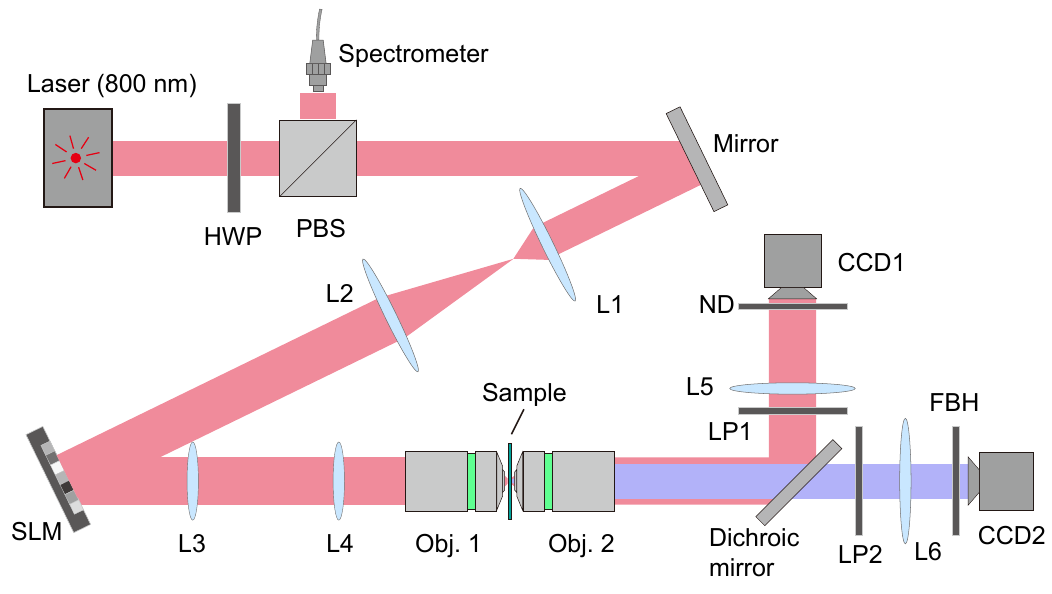}
  } 
  \caption{\noindent\textbf{Experimental setup}. HWP: half wave plate; PBS: polarizing beam splitter; L: lens; SLM: spatial light modulator; Obj.: objective; LP: linear polarizer; FBH: hard-coated bandpass filter;  ND: neutral density filter; CCD: charge-coupled device (camera).}
 \label{FigureS2}
\end{figure*} 

\begin{figure*}[!t]
  \renewcommand{\figurename}{Supplementary Figure}
  \centering{
  \includegraphics[width = 0.5\linewidth]{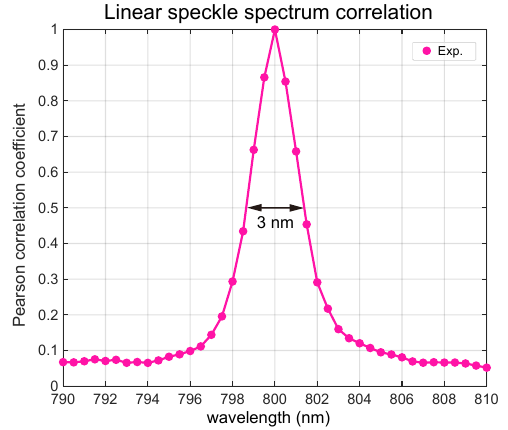}
  } 
  \caption{\noindent\textbf{Linear speckle spectrum correlation.} We switch the femtosecond laser to monochromatic operation to measure the correlation bandwidth of the linear speckle. We vary the FH wavelength from 790 nm to 800 nm and calculate the speckle correlation against the speckle measured at 800 nm. The correlation bandwidth is estimated as around 3 nm (FWHM, full width at half maximum). As such, the bandwidth of the pulsed mode of the laser used in the experiment is just a few times of the correlation bandwidth of the medium at FH wavelength.}
 \label{FigureS3}
\end{figure*} 

\begin{figure*}[!t]
  \renewcommand{\figurename}{Supplementary Figure}
  \centering{
  \includegraphics[width = 0.65\linewidth]{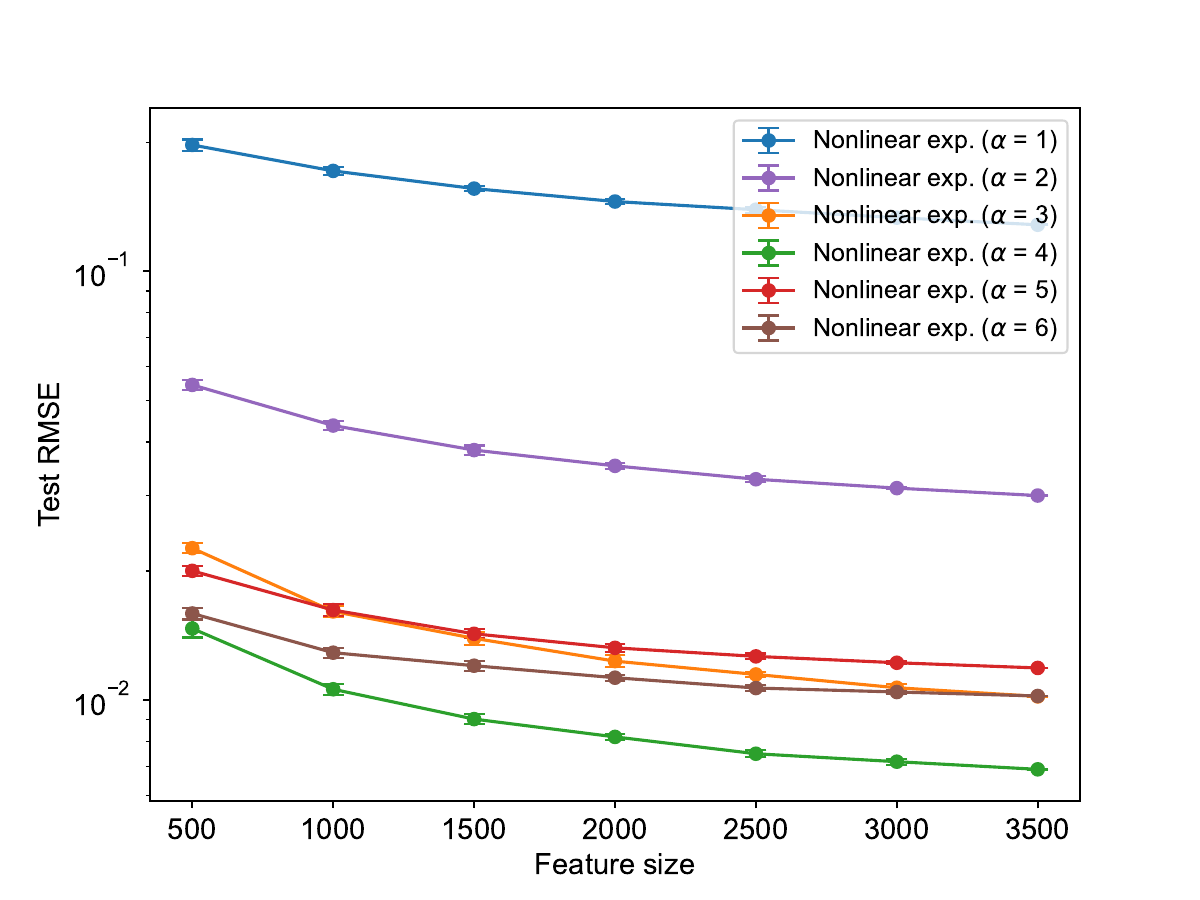}
  } 
  \caption{\noindent\textbf{Sinc function interpolation with different phase encoding ranges.} The test RMSEs based on the experimental nonlinear features at varying feature sizes, where $\alpha$ indicates the phase encoding range of $[0,\alpha\pi]$.}
 \label{FigureS4}
\end{figure*} 

\begin{figure*}[!t]
  \renewcommand{\figurename}{Supplementary Figure}
  \centering{
  \includegraphics[width = 1.0\linewidth]{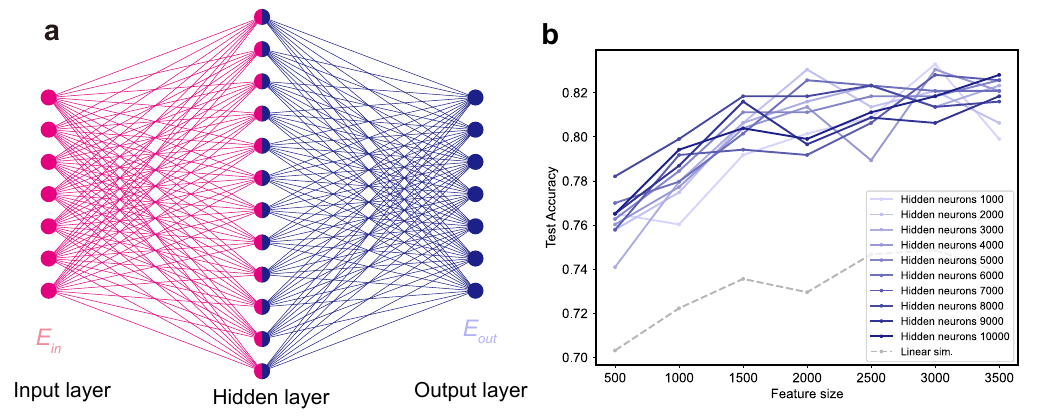}
  } 
  \caption{\noindent\textbf{Simulation of SLD image classification with a simplified nonlinear forward model.} \textbf{a,} A two-layer random network with square nonlinearity in the hidden layer as a simplified nonlinear forward model ignoring the second-order nonlinear polarization of LN crystals. \textbf{b,} The simulated test accuracies based on two-layer neural networks with varying numbers of hidden neurons. Blue solid lines: simulated nonlinear results; Gray dashed line: simulated linear random projection (same as that of Fig. 2d in the main text).}
 \label{FigureS5}
\end{figure*}

\begin{figure*}[!t]
  \renewcommand{\figurename}{Supplementary Figure}
  \centering{
  \includegraphics[width = 1.0\linewidth]{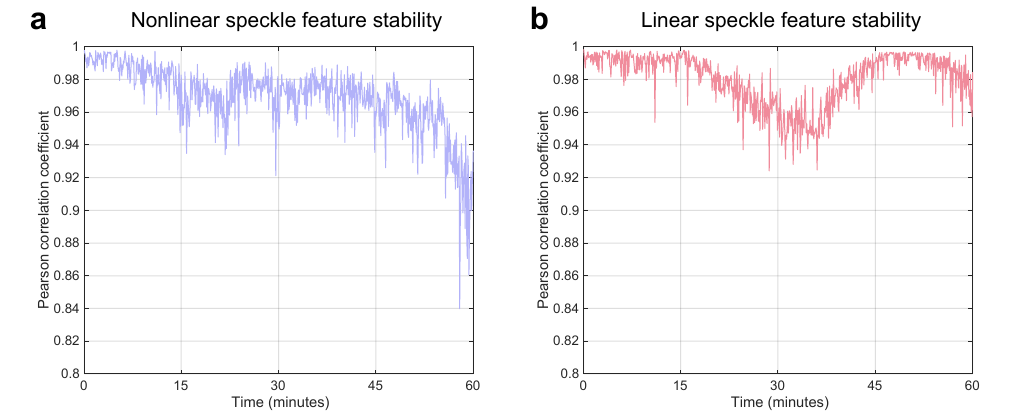}
  } 
  \caption{\noindent\textbf{Experimental system stability.} \textbf{a,} Nonlinear Pearson correlation coefficient evolution over an one-hour duration. \textbf{b,} Linear Pearson correlation coefficient evolution over an one-hour duration. }
 \label{FigureS6}
\end{figure*} 

\clearpage

\begin{table}[hbt!]
\begin{threeparttable}
\caption{Summary of experimental results
\label{Table S1}
}
\begin{center}
\scriptsize\centering
\begin{tabular}{ m{2cm}<{\centering} m{2.4cm}<{\centering} m{2.4cm}<{\centering} m{2.8cm}<{\centering} m{2cm}<{\centering} m{2.4cm}<{\centering} }
\hline
\textbf{Dataset} & \textbf{Task description} & \textbf{Training/test size} & \textbf{Sample dimension} & \textbf{Macro-pixel size} & \textbf{Average optical nonlinearity gain}  \\
\hline
SLD & Image classification (10 categories) & 1,649/413 & 64$\times$64 & 5$\times$5 & 5.4\% $\sim$ 10.9\% \\
ASL & Image classification (24 categories) & 27,455/7,172 & 28$\times$28 & 10$\times$10 & 4.0\% $\sim$ 5.0\% \\
CIFAR-10 & Image classification (10 categories) & 50,000/10,000 & 32$\times$32$\times$3 & 5$\times$5 & 0.7\% $\sim$ 2.8\% \\
Sinc function & Interpolate the sinc function (univariate regression) & 2,400/600 & 1 & 10$\times$10 & -0.055 $\sim$ -0.043\tnote{a}  \\
UCI yacht hydrodynamics & Predict the residuary resistance of sailing yachts (multivariate regression) & 214/94 & 6 & 150$\times$100 & 0.042 $\sim$ 0.046\tnote{b} \\
Concrete compressive strength & Predict the concrete compressive strength (multivariate regression) & 927/103 & 8 & 150$\times$75 & 0.078 $\sim$ 0.087\tnote{b} \\
SBM-generated graph dataset & Graph classification (2 categories) & 400/100 & (60, 254) $\sim$ (60, 360), average (60, 300.35)\tnote{c} & 5$\times$5 & 2.1\% $\sim$ 3.7\%\tnote{d}\\
Reddit-binary graph dataset & Graph classification (2 categories) & 1,800/200 &  (6, 4) $\sim$ (3,782, 4,071), average (431.77, 500.24)\tnote{c} & 5$\times$5 & 0.7\% $\sim$ 3.4\%\\
MNIST & Image classification (10 categories) & 60,000/10,000 & 28$\times$28  & 10$\times$10 & 0.9\% $\sim$ 1.1\%\\
FashionMNIST & Image classification (10 categories) & 60,000/10,000 & 28$\times$28  & 10$\times$10 & -0.2\% $\sim$ 0.9\%\\
STL-10 & Image classification (10 categories) & 8,000/5,000 & 96$\times$96$\times$3 & 2$\times$2 & 2.4\% $\sim$ 3.5\%\\

\hline
\end{tabular}
\begin{tablenotes}
% Note: 
    \footnotesize
    \item[a] The metric used is RMSE. The '-' sign indicates a decrease in performance.
    \item[b] The metric used is $R^2$. 
    \item[c] The graph samples are represented with the notation (\textit{nodes, edges}). For example, (60, 254) $\sim$ (60, 360), and (60, 300.35) of SBM-generared graphs denote the smallest, largest, and average number of nodes and edges in the graphs.
    \item[d] SBM-generated graphs with an inter-class similarity of $k = 1.4$.
\end{tablenotes}

\end{center}
\end{threeparttable}
\end{table}

\renewcommand{\bibpreamble}{
$^\ast$These authors contributed equally to this work.\\
$^\dagger${Corresponding author: \textcolor{magenta}{jianqi.hu@lkb.ens.fr}}\\
$^\ddag${Corresponding author: \textcolor{magenta}{sylvain.gigan@lkb.ens.fr}}
}

\bibliographystyle{naturemag}
\bibliography{ref}